\documentclass[12pt]{iopart}

\usepackage{epsfig,color,colordvi, ulem}

\usepackage{iopams}

\def\be{\begin{equation}}
\def\ee{\end{equation}}
\def\bee{\begin{eqnarray}}
\def\eee{\end{eqnarray}}

\def\kb{k_{\rm B}}
\def\tilde{\widetilde}
\def\L{{\cal L}}
\def\U{{\cal U}}
\def\M{{\cal M}}

\def\halb{\mbox{$\frac{1}{2}$}}
\def\ihalb{\mbox{$\frac{i}{2}$}}

\begin{document}

\title{Symmetry reduction for tunneling defects due to strong couplings to phonons}

\author{P. Nalbach$^1$ and M. Schechter$^2$}
\address{
$^1$Westf\"alische Hochschule, M\"unsterstr. 265, 46397 Bocholt, Germany\\
$^2$Department of Physics, Ben Gurion University of the Negev, Beer Sheva
84105, Israel}
\date{\today}

\begin{abstract}
Tunneling two-level systems are ubiquitous in amorphous solids, and form a major source of noise in systems such as nano-mechanical oscillators, single electron transistors, and superconducting qubits. Occurance of defect tunneling despite their coupling to phonons is viewed as a hallmark of weak defect-phonon coupling. This is since strong coupling to phonons results in significant phonon dressing and suppresses tunneling in two-level tunneling defects effectively.
Here we determine the dynamics of a tunnelling defect in a crystal strongly coupled to phonons incorporating the full 3D geometry in our description. We find that inversion symmetric tunnelling is not  dressed by phonons whereas other tunnelling pathways are dressed by phonons and, thus, are suppressed by strong defect-phonon coupling. We provide the linear acoustic and dielectric response functions for a tunnelling defect in a crystal for strong defect-phonon coupling. This allows direct experimental determination of the defect-phonon coupling.
The singling out of inversion-symmetric tunneling states in single tunneling defects is complementary to their dominance of the low energy excitations in strongly disordered solids as a result of inter-defect interactions for large defect concentrations. This suggests that inversion symmetric two-level systems play a unique role in the low energy properties of disordered solids.

\end{abstract}

\maketitle

\section{Introduction}

The generic existence of tunneling two-level systems (TLSs) in disordered and amorphous systems has been postulated more than four decades ago \cite{AHV72,Phi72} to explain the low temperature universality in amorphous solids \cite{ZP71,FA86,HR86,Phi87, Yu88, PLT02, Ramos2014}. Recently, TLSs have attracted strong renewed interest since they are a dominant dephasing source in various nano devices, and in particular in superconducting quantum devices. TLSs are found in surface oxides of thin-film circuits electrodes \cite{Gao08}, at disordered interfaces \cite{Quint14} and in the tunnel barrier of Josephson junctions \cite{SRW04}. They possess electric as well as elastic moments with which they couple to their environment and create noise for nearby electronic devices. The physical origin of the tunneling defects in these applications is still as little understood as in the disordered and amorphous bulk systems. Surprisingly, superconducting qubits suffer dominantly from single, dominant TLSs \cite{Bar14, Mue15, LBM+16}.

Individual tunneling defects have been observed in the form of substitutional defect ions in alkali halide crystals where they lead to particular low temperature properties \cite{Nara}. Individual tunneling defects are only observed at lowest defect concentrations, i.e. a few to tens of ppm. At defect concentrations above 100 ppm one typically faces a complicated many body problem \cite{Aloisbuch, Lud04} of interacting defects. Over a wide range of very large concentrations in the percent regime these systems then show glassy behaviour\cite{HKL90}, i.e. the low temperature universal properties of glasses, as revealed in the nearly linear specific heat and the particular characteristics related to phonon attenuation\cite{YKMP86}.

The individual tunneling defects in crystals are not simple TLSs. In contrast, the substitutional defect atom tunnels between 6, 8 or 12 potential minima due to the host crystals cubic symmetry \cite{Gomez}. Nevertheless, thermal and dielectric properties of tunneling states in crystals can typically be studied using a simplified two-states model \cite{Aloisbuch,Nal97}. In contrast, the acoustic response shows a more complex behaviour reflecting the defect geometry and the according tensor character of the elastic moment \cite{Nal01}.

It was shown that, despite the multi-level nature of the defects in disordered lattices, tunneling states with multi degeneracy at small disorder, unfold upon strong disorder, i.e. at high defect concentrations, to pairs of states related to each other by local inversion \cite{SS13}. These pairs of states possess a small energy bias within each inversion symmetry pair, and large bias in between pairs\cite{SS13,CBS14}. This mechanism leads to dominance of inversion symmetric two-level systems in the density of low energy excitations of strongly disordered solids. Here, we show that strong defect-phonon coupling suppresses tunneling except along inversion symmetric pathways. Thus, strong defect phonon coupling provides an additional mechanism, through its symmetry dependent attenuation of the tunneling amplitude, for the preference of locally inversion symmetric two level systems as the entities constituting the low energy tunneling systems in disordered solids.

The typically reported defect-phonon coupling strength is weak to intermediate. Most experimental data, however, is acquired in samples with defect concentrations $n\gtrsim 100$ ppm, where many body effects complicate the situation. From a first-principle argumentation one actually expects for tunneling defects with no special symmetry, rather strong defect-phonon couplings \cite{SS13,Gai11}.

In this paper we determine the dynamics of a tunnelling defect with cubic symmetry strongly coupled to phonons, using a model which incorporates the geometric structure of the defects and goes beyond the two-state approximation. We consider a single defect in an otherwise perfect crystal. In contrast to the two-state approximation we find that strong defect-phonon coupling does not suppress tunnelling. It reduces the cubic symmetry of the defect effectively to inversion symmetric states, between which tunneling remains nearly unaffected by the interaction. This complements the paradigm that a defect strongly coupled to phonons cannot exhibit tunnelling. Furthermore, we provide the linear acoustic and dielectric response functions for the tunnelling defects strongly coupled to phonons which allows direct experimental determination of the defect-phonon coupling. These findings allow to improve substantially the characterization of tunnelling defects in a crystal which is necessary to unravel the low temperature properties of crystals with higher defect concentrations, which show glassy behaviour. This might further help to unravel the nature of tunnelling defects in amorphous solids, and ultimately to the reduction of two-level systems' caused dephasing in superconducting qubits\cite{SRW04,SSMM05,MJM05} and nanomechanical oscillators\cite{Nems08, Nems09, Nems10}, a significant obstacle in the quest of quantum computing and quantum metrology.

We organize the paper as follows: In section II we introduce substitutional [111] tunnelling defects and the model within which their tunneling dynamics is discussed. In section III we discuss the coupling of these defects to phonons. Strong defect-phonon coupling is treated within a polaron approach which we introduce in section IV. The dielectric and acoustic response is presented in section V. Finally, in section VI, we discuss the experimental relevance of our results for [111] defect systems like KCl:Li and KCl:CN, as well as for TLSs in superconducting qubits. We then end with conclusions.

\section{The [111] defect}

\begin{figure}
\begin{center}
\includegraphics[width=10cm]{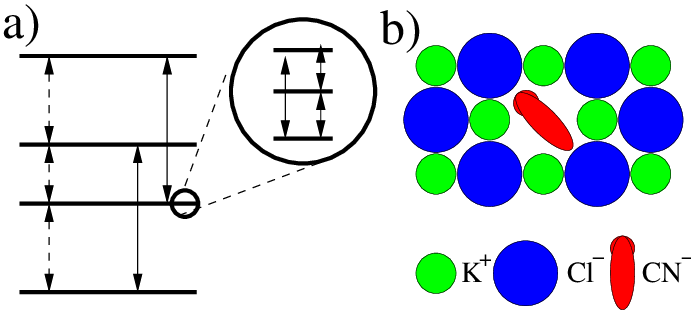}
\end{center}
\caption{\label{spectrum} a) Energy spectrum of a [111]-impurity without
coupling to phonons. Dashed arrows indicate the allowed dipolar transitions,
and full arrows the quadrupolar ones. b) 2D illustration for a CN impurity
in KCl forming a [111]-defect}
\end{figure}

Typical defect systems are KCl doped with Li, OH or CN molecules. The potential energy landscape in which the defect ion moves is given by the host crystal and therefore reflects its symmetry which for most alkali halide crystals is cubic (e.g. the fcc-structure of potassium chloride). This results for KCl doped with Li or CN molecules in eight potential wells in [111]-directions \cite{Bey75}, i.e. at the corners of a small cube with side length $d$. The dopant OH has 6 potential wells in [100]-directions \cite{Kapp74, Lud00}. At low temperatures thermally activated crossing over potential barriers is inhibited for the defect ions and quantum tunnelling remains and typically leads to a ground state splitting of about 1 Kelvin. Since the number density of such tunnelling states exceeds, even at low concentration, that of small-frequency phonon modes of the host crystal, the impurities govern the low-temperature properties of the material. We focus on defects with 8 minima in the [111]-direction (see Fig.\ \ref{spectrum}b for a 2D illustration) and refer to these defect
systems as [111]-defects in the following.

To model a [111] defect at low temperatures \cite{Nara, Aloisbuch, Gomez} one restricts the Hilbert space to the 8 localized impurity positions with corresponding localized states. For a lithium impurity, the off-center positions ${\bf r}$ form a cube of side-length $d$, whereas for cyanide impurities ${\bf r}$ indicates the orientation of the cigar-shaped polar molecule. A [111] impurity can be described as the product of three two-state variables \cite{Nal97}. We adopt the shorthand notation for its quantum operators
\begin{equation} \label{e8}
A_{\alpha\beta\gamma}
= \sigma_\alpha^1\otimes\sigma_\beta^2\otimes\sigma_\gamma^3 ,
\end{equation}
where $i=1,2,3$ label the crystal axes and Greek indices $\alpha=0,x,y,z$ label the usual Pauli matrices with $\alpha=0$ for the identity operator. In a two-state tunneling defect the operator $\sigma_z$ reflects the {\it position} since it distinguishes between {\it left} and {\it right}. At the same time $\sigma_x$ reflects {\it tunneling} by allowing transitions between the localized quantum states. In the same way, for the [111] defect the {\it position} operator is given as ${\bf r}= (d/2){\bf e}$ with
\be
{\bf e} = \left( \begin{array}{c} A_{z00} \\ A_{0z0} \\ A_{00z} \end{array}  \right) .
\ee

A [111]-impurity can tunnel via three different paths between its eight identical potential minima: (i) along the edges of the cube with corresponding tunnel coupling $\Delta_k$, (ii) along a face diagonal with tunnel coupling $\Delta_f$ and (iii) along a space diagonal with tunnel coupling $\Delta_r$. The corresponding Hamiltonian is
\be\label{Hsys}
H_S = -\frac{\Delta_k}{2} \left( A_{x00} + A_{0x0} + A_{00x} \right) - \frac{\Delta_f}{2} \left( A_{xx0} + A_{x0x} + A_{0xx} \right)  - \frac{\Delta_r}{2}  A_{xxx} . \quad
\ee
Note that, here, the tunnel couplings are model parameters. Since tunnel couplings depend exponentially on the length of their respective tunnelling path \cite{Aloisbuch}, one typically concludes \cite{Gomez} that $\Delta_k\gg\Delta_f\gg\Delta_r$. Then, the energy spectrum (shown in Fig. \ref{spectrum}a) has four almost equidistant energy levels with splitting $\Delta_k$. The upper and lower ones are single quantum states, whereas the middle levels are threefold degenerate. Neglecting face and space diagonal tunnelling completely, the problem factorizes into three two-level systems (TLS), one for each tunneling direction, \cite{Nara,Nal97} with energy splitting $\Delta_k$. This justifies two-state approximations for thermal and dielectric properties but not for the more complex acoustic response due to the tensor character of the elastic moment \cite{Nal01}.

Electic fields couple to the dipole ${\bf p} = p_0{\bf e}$ of the defect and cause transitions between adjacent levels as highlighted by the dashed arrows in the spectrum (see Fig. \ref{spectrum}a).

For a lattice distortion that varies sufficiently slowly in space, the interaction potential is given by the term that is linear in the elastic strain $\epsilon_{jl}({\bf r})$,
\begin{equation}\label{e9b}
W({\bf r}) \,=\, -\sum_{jl} Q_{jl}({\bf r})\epsilon_{jl}({\bf r}) \; .
\end{equation}
Equation (\ref{e9b}) is the lowest-order term \cite{foot5} of a multipole expansion with the elastic quadrupole operator \cite{Aloisbuch,Nal01}
\begin{equation} \label{e10}
Q_{jl} = \gamma\,e_{j}e_{l} ( 1- \delta_{jl}),
\end{equation}
with, for example, $Q_{xy}=\gamma A_{zz0}$, the elastic coupling energy $\gamma$ and the components $e_j$ and $e_l$ of ${\bf e}$. Elastic perturbations induce two types of transitions as indicated by the solid arrows in Fig. \ref{spectrum}a, namely between ground and second excited and between first and third excited states, but also between the degenerate states of the first excited states and between the degenerate states of the second excited states.

\section{Defect-Phonon coupling}

The elastic strain at the defect position ${\bf r}$ due to phonons \cite{Aloisbuch,Nal01} is determined as spatial derivative $\epsilon_{jl}({\bf r})=\partial_{x_j}u_l({\bf r})$ of the displacement amplitude \cite{foot1} ${\bf u}({\bf r}) = \sum_{{\bf k},\alpha} i{\boldsymbol \xi}_{{\bf k}\alpha} e^{i\bf kr} q_{{\bf k}\alpha}$, with polarisation vector ${\boldsymbol \xi}_{{\bf k}\alpha}$ for phonon mode with frequency $\omega_{{\bf k}\alpha}$, wave vector ${\bf k}$ and polarisation $\alpha$ for longitudinal and transverse phonon branches and displacement operator $q_{{\bf k}\alpha}$ of mode ${\bf k}$.
Within our [111]-defect model the position of the defect ion ${\bf r}$ is a discrete quantum operator. We assume Debye phonons with linear dispersion up to the Debye frequency $\omega_D$ which corresponds to a wavelength $\lambda_D\simeq a$ with lattice constant $a$. Since $a>d$ and thus $k_jd < 1$ for all modes, we expand $e^{i{\bf kr}}$ and neglect all terms beyond linear order in $k_jd$ for $j=1,2,3$. Inserting this in Eq. (\ref{e9b}) results in a defect-phonon coupling Hamiltonian
\be \label{defphoncoup}
W_{dp} = W_s + W_{w,1}+W_{w,3}
\ee
with
\bee \label{ws}
W_s &=& \sum_{{\bf k},\alpha} ( A_{zz0} \lambda_{s,xy}({{\bf k},\alpha}) +A_{z0z}  \lambda_{s,xz}({{\bf k},\alpha}) +A_{0zz}\lambda_{s,yz}({{\bf k},\alpha}) ) \cdot q_{{\bf k}\alpha} \\ \label{ww3}
W_{w,3} &=& A_{zzz} \sum_{{\bf k},\alpha} \lambda_{w,3}({{\bf k},\alpha})  q_{{\bf k}\alpha} \\ \label{ww1}
W_{w,1} &=& \sum_{{\bf k},\alpha} (  A_{z00} \lambda_{w,1,x}({{\bf k},\alpha}) +A_{0z0} \lambda_{w,1,y}({{\bf k},\alpha}) +A_{00z} \lambda_{w,1,z}({{\bf k},\alpha}) ) \cdot q_{{\bf k}\alpha} \quad
\eee
therein defining the defect-phonon coupling constants $\lambda_\phi ({{\bf k},\alpha})$.

The full Hamiltonian of the defect plus phonons becomes (with momentum $p_{{\bf k}\alpha}$ of the phonon mode)
\be\label{hamges}
 H = H_S + W_s + W_{w,1} +W_{w,3} + \halb\sum_{{\bf k}\alpha} \left( p^2_{{\bf k}\alpha}
+ \omega^2_{{\bf k}\alpha} q_{{\bf k}\alpha}^2 \right).
\ee
In order to determine the effect of the phonons onto the dynamics of the defect we follow a standard system-bath approach \cite{WeissBuch, SpiBoLe1987}. Each operator $A_{xyz}$ in Eq. (\ref{ws}), (\ref{ww3}) and (\ref{ww1}) results in a bath spectral function:
\be\label{spectra}
J_\phi (\omega) = \alpha_\phi \omega^{x_\phi} e^{-\frac{\omega}{\omega_D}}
\ee
with equal $\alpha_s$ and $x_s=3$ for $\phi=(s,xy)$, $(s,xz)$, $(s,yz)$ and
$\alpha_{w,3}$ and $x_{w,3}=5$ for $\phi=(w,3)$ and $\alpha_{w,1}$ and
$x_{w,1}=5$ for $\phi=(w,1,x)$, $(w,1,y)$ and $(w,1,z)$ (see appendix \ref{app1} for details). Since $J_{w,3}(\omega)\simeq J_{w,1}(\omega) \ll J_{s}(\omega)$, $W_s$ is the dominant defect-phonon coupling.

\section{Polaron transformation}

The dominant contribution to the coupling between tunnel defect and phonons is $W_s$ and, potentially, it is a {\it strong} defect phonon coupling. Strong system-bath coupling can successfully be treated employing the non-interacting blip approximation (NIBA) as introduced by Leggett et al. \cite{SpiBoLe1987}. Identical results are obtained when a Polaron transformation with a subsequent lowest order Born-Markov approximation \cite{Dekker, Wue98} is used. We follow the second route.

Defining shift operators
\[
F_{{\bf k}\alpha} = \left( A_{0zz} f_{yz,{\bf k}\alpha}
+A_{z0z} f_{xz,{\bf k}\alpha} +A_{zz0} f_{xy,{\bf k}\alpha} \right)
\]
with $f_{jl,{\bf k}\alpha}=\lambda_{s,jl}({\bf k},\alpha)/\omega^2_{{\bf k}\alpha}$ and the according Polaron transformation
\be\label{polartrafo}
 T=\exp\left(-\ihalb\sum_{{\bf k}\alpha} F_{{\bf k}\alpha}p_{{\bf k}\alpha}\right) ,
\ee
and defining $\tilde{H} = T^\dag\cdot H\cdot T$, leads to (see appendix \ref{app2} for details)
\be\label{hamgestraf}
 \tilde{H} = \tilde{H}_S + \tilde{W}_s + W_{w,1}+W_{w,3} + \halb\sum_{{\bf k}\alpha}
\left( p^2_{{\bf k}\alpha}
+ \omega^2_{{\bf k}\alpha} q_{{\bf k}\alpha}^2 \right) + H_F\, ,
\ee
with $H_F = -\halb\sum_{{\bf k}\alpha}\omega^2_{{\bf k}\alpha} F_{{\bf k}\alpha}^2 $. An easy but tedious calculation shows that $H_F$ is a constant shift of the zero point energy only. The Polaron transformation only shifts the defect Hamiltonian (\ref{Hsys}) and the dominant defect-phonon coupling (\ref{ws}) whereas the subdominant defect-phonon couplings (\ref{ww3}) and (\ref{ww1}) are unmodified.

Supprisingly, the Polaron transformation leaves space diagonal tunnelling, which inverts the impurity position, $T^\dag\cdot A_{xxx} \cdot T =  A_{xxx}$ unmodified. In contrast, edge as well as face diagonal tunnelling are {\it dressed} leading to
\be\label{hsystraf}
\tilde{H}_S = -\frac{\Delta_k W}{2} \left( A_{x00} + A_{0x0} + A_{00x} \right)  - \frac{\Delta_f W}{2} \left( A_{xx0} + A_{x0x} + A_{0xx} \right)  - \frac{\Delta_r}{2}  A_{xxx}
\ee
with the Debye-Waller {\bf factor} (see appendix \ref{app2} for details)
\be
W = \exp\left(-(\alpha_s/\pi)\int_0^\infty d\omega \omega \coth(\beta\omega/2)
e^{-\omega/\omega_D}  \right) .
\ee
For low temperatures, i.e. $\beta^{-1}=\kb T \ll \hbar\omega_D $ which holds for typical experimental temperatures of about 1 Kelvin, $W\lesssim \exp(-(\alpha_s/\pi)\omega_D^2)$. Thus, for strong system-bath coupling, i.e. $\alpha_s\omega_D^2\gg 1$, one finds $W\ll 1$.

For a two-state system strongly coupled to phonons tunnelling is strongly suppressed by the Debye-Waller factor. Surprisingly, this does not hold for the [111] defect. Whereas edge and face diagonal tunnelling is suppressed, space diagonal (inversion symmetric) tunnelling is not influenced by the dominant defect-phonon coupling due to symmetry. {\it Thus, strong defect-phonon coupling does not suppress tunnelling in a [111] defect.}

Face and space diagonal tunnelling are expected to be subdominant compared to edge tunnelling due to the longer geometric tunnelling path. Phonon renormalization of edge and face diagonal tunnelling competes now with the geometrical suppression of the face and space diagonal tunnelling. For weak defect-phonon coupling $W\simeq 1$, where edge tunnelling dominates, an energy spectrum as depicted in Fig. \ref{spectrum}a results. The dynamic dielectric and acoustic response for this case have been discussed previously \cite{Nal01}.

\begin{figure}
\begin{center}
\includegraphics[width=8cm]{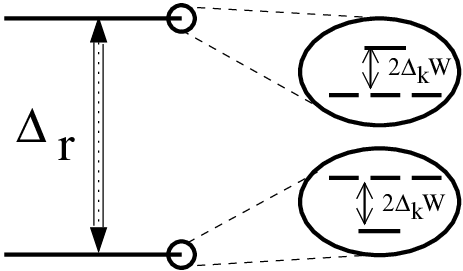}
\end{center}
\caption{\label{spectrumII} Energy spectrum of a [111]-impurity for very strong
defect-phonon coupling. Double lined arrow indicates the allowed dielectric
transitions, and single lined arrow {\bf indicates the allowed acoustic transitions}.}
\end{figure}

In contrast, when the phonon renormalization dominates, i.e. $\Delta_r\gg\Delta_kW$, space diagonal (inversion symmetric) tunnelling will dominate leading to an energy spectrum with two fourfold degenerate states with splitting $\Delta_r$ (as depicted in Fig. \ref{spectrumII}). The lower (upper) states have inversion (anti) symmetry. Importantly, only dielectric transitions are allowed between the two fourfold degenerate states. The suppressed edge tunnelling leads to a small splitting of the fourfold degeneracy into two states, one singlet and one triplet, between which acoustic transitions are allowed. As face diagonal tunnelling is subdominant in both cases, we set $\Delta_f=0$ in the following, for simplicity.

\section{Dynamic response}

Next, we determine the response of the defect to applied dielectric and acoustic
fields. The treatment of the remaining defect-phonon couplings $\tilde{W}_s + W_{w,1}+W_{w,3}$ in the Hamiltonian (\ref{hamgestraf}) is done within a standard open quantum system framework \cite{WeissBuch} employing a resumed perturbative treatment in a super-operator formulation (RESPET) \cite{Horner, Nal02, Nal10a, Nal14}. Details are outlined in appendix \ref{app3}.

RESPET allows to determine the correlation function $C_{AB}(t) = \halb\langle A(t)B+BA(t) \rangle$ for defect operators $A$ and $B$ under the influence of phonons. Explicitly, we determine its Laplace transform $C_{AB}(z)$.
The spectrum $C_{AB}''(\omega)$ is the imaginary part of the Laplace transform continued to the real axis. It is connected to the spectrum  $\chi_{AB}''(\omega)$ of the linear response function $\chi_{AB}(t)=\langle A(t)B-BA(t) \rangle$ via the fluctuation-dissipation-theorem
\be
\chi''_{AB}(\omega) = (2/\hbar)\tanh(\beta\omega/2) C''_{AB}(\omega) .
\ee
Absorption or dielectric / acoustic loss are proportional to the spectra $\chi''_{AB}(\omega)$ whereas change of dielectric constant or speed of sound are proportional to the real part of the susceptibility $\chi_{AB}'(\omega)$ which can be determined using the Kramers-Kronig relation.

\subsection{Dielectric response}

Due to the cubic symmetry dielectric responses for electric fields polarized in any direction are identical. We, thus, focus on the $x$ direction and determine the correlation function of the dipole operator $p_x$ (neglecting damping)
\be\label{eq16}
C_{p_xp_x}(z) = \frac{n_1(T)z}{z^2-\delta^2} + \frac{n_2(T)z}{z^2-\Delta^2} \, .
\ee
Here
\[
n_2(T)=\sum_{j=1}^4 e^{-\beta E_j}/Z(T) \:\,\mbox{and}\:\, n_1(T)=\sum_{j=2}^3 2
e^{-\beta E_j}/Z(T)
\]
depend on the thermal occupations of the involved states,
with $Z(T)=e^{-\beta E_1}+3e^{-\beta E_2}+3e^{-\beta E_3}+e^{-\beta E_4}$, and energies $E_{1}=- (\Delta_r+3\Delta_k W)/2$, $E_4=-E_1$, $E_{2}= -(\Delta_r-\Delta_k W)/2$ and $E_3=-E_2$. We also define $\delta=\Delta_r-\Delta_kW=E_3-E_1=E_4-E_2$ and $\Delta=\Delta_r+\Delta_kW=E_3-E_2$.

Whereas bias asymmetric two-level systems (i.e. two-level systems having non-zero bias energy) exhibit resonant and relaxational contributions to the response \cite{Phi87}, bias symmetric two-level systems only have a resonant contribution. We treat a [111]-defect without disorder which might cause bias asymmetry and observe accordingly resonant contributions to the response only.

At strong coupling, i.e. $\Delta_r\gg\Delta_kW$, Eq.(\ref{eq16}) simplifies to $C_{p_xp_x}(z)=z/(z^2-\Delta_r^2)$ whereas at weak coupling, i.e. $\Delta_r\ll\Delta_kW$ and $W\simeq 1$, we obtain formally the same function with $\Delta_r\rightarrow\Delta_k$. Since the tunnel couplings are not known, a priori, measuring the dielectric response provides the energy splitting but cannot differentiate between strong or weak defect-phonon coupling. However, for intermediate couplings the situation changes. The dielectric response becomes at low frequencies $\hbar\omega\ll\kb T, \Delta,\delta$
\be
 \chi'_{p_xp_x}(T) = \frac{2 n_2(T)}{\Delta}\tanh\left(\frac{\Delta}{2 \kb T}\right)
+ \frac{2n_1(T)}{\delta}\tanh\left(\frac{\delta}{2\kb T}\right) .
\ee
\begin{figure}
\begin{center}
\includegraphics[width=9.5cm]{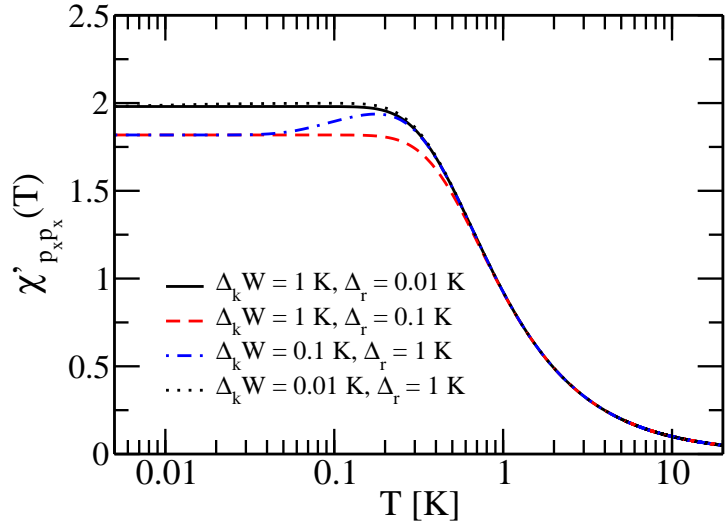}
\end{center}
\caption{\label{chip} Low frequency ($\hbar\omega\ll\kb T$) dielectric response $\chi'_{p_xp_x}(T)$ is plotted versus temperature for several effective tunnel couplings reflecting cases from weak to strong defect-phonon coupling.}
\end{figure}
No significant contribution to the spectrum $\chi''_{p_xp_x}(T)$ is observed in this regime. Fig. \ref{chip} plots the $\chi'_{p_xp_x}(T)$ versus temperature. At weak defect-phonon coupling with $\Delta_r\ll\Delta_kW$ (black full line) one observes the expected $tanh$ behaviour known for two-level systems. With increasing $\Delta_r$ (red dashed line) the plateau value at lowest frequencies diminishes but otherwise the qualitative behaviour is unchanged. Once $\Delta_r$ dominates at strong but not very strong defect-phonon coupling a hump at about $T\simeq 0.2$ evolves (dash-dotted blue line).
This hump results from the temperature dependence of $n_1(T)$ in Eq.(\ref{eq16}). It is not a relaxational feature commonly observed in two-level systems \cite{Phi87} since the according strong frequency dependence and contributions to the spectrum $\chi''_{p_xp_x}(T)$ are missing here \cite{foot4}.
At very strong defect-phonon coupling with $\Delta_r\gg\Delta_kW$ (dotted black line) we observe again the same $tanh$ behaviour as at weak coupling. Damping does not alter the above results since the super-Ohmic spectra (\ref{spectra}) ensure that all damping rates $\Gamma\ll \Delta, \delta$ and frequencies $\omega$ of external fields applied experimentally are $\omega\ll \Delta, \delta$ as well.

\subsection{Acoustic response}

\begin{figure}
\begin{center}
\includegraphics[width=9.5cm]{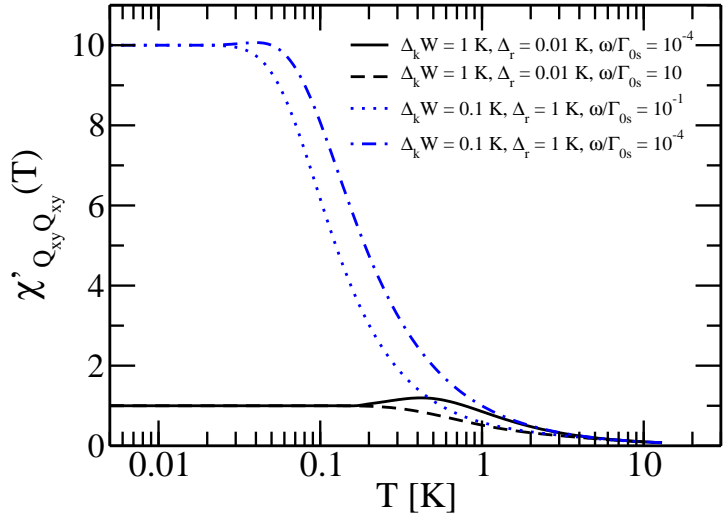}
\end{center}
\caption{\label{chiq} Low frequency ($\hbar\omega\ll\kb T$) acoustic response $\chi'_{Q_{xy}Q_{xy}}(T)$ is plotted versus temperature for weak, i.e. $\Delta_kW\gg\Delta_r$, and strong defect-phonon coupling $\Delta_kW\ll\Delta_r$. Each case is shown for two different experimental frequencies reflecting situations with fully and partially suppressed relaxational contributions.}
\end{figure}

We focus on the response to the elastic operator $Q_{xy}$ and observe (neglecting damping at first)
\be\label{cq1}
C_{Q_{xy}Q_{xy}}(z) = \frac{n_2(T) z}{z^2-(2\Delta_kW)^2} + \frac{n_1(T)}{z}
\ee
which exhibits a relaxational contribution (second term on r.h.s in Eq.(\ref{cq1})) and a resonant contribution (first term on r.h.s in Eq.(\ref{cq1})) which is governed for any defect-phonon coupling by the edge tunnel coupling $\Delta_kW$.
Note that unlike in the dielectric response we observe here relaxational contributions despite treating a bias symmetric tunneling defect in contrast to two-level system behaviour. These relaxational contributions are a result of the degeneracies of the spectrum \cite{Nal01}.

Remarkably, in Eq.(\ref{cq1}) for any defect-phonon coupling the tunnel coupling $2\Delta_kW$ governs the acoustic response. Thus, at strong defect-phonon coupling, the dielectric response is governed by $\Delta_r$ whereas the resonant contribution of the acoustic response is governed by $2\Delta_kW$. Measuring both and observing different relevant energies, then, points towards the strong coupling scenario.

Next, we take damping into account and determine the acoustic response for small experimental frequencies $\hbar\omega\ll\kb T$. As long as $\Delta_kW\gg\hbar\omega$, we obtain (see appedix C and references therin for details)
\be\label{chiqeq}
\chi'_{Q_{xy}Q_{xy}}(\omega) = \beta \frac{2e^{-\beta E_2}}{Z(T)}
     \frac{\Gamma_b^2}{\omega^2+\Gamma_b^2}
     \, + \beta \frac{2e^{-\beta E_3}}{Z(T)}
     \frac{\Gamma_c^2}{\omega^2+\Gamma_c^2}\,  + \frac{n_2(T)}{\Delta_kW}\tanh\left(\frac{\Delta_kW}{\kb T}\right)
\ee
with rates \cite{foot3}
\bee
\Gamma_b &=& 2\pi J_{s}(2\Delta_kW) \left( 1 + n_B(2\Delta_kW) \right)  + 2\pi\left[ J_{w,3}(\Delta_r)+3J_{w,1}(\Delta_r) \right] n_B(\Delta_r)\nonumber\\
\Gamma_c &=& 2\pi J_{s}(2\Delta_kW) n_B(2\Delta_kW)  + 2\pi \left[ J_{w,3}(\Delta_r)+3J_{w,1}(\Delta_r) \right] \left( 1 + n_B(\Delta_r) \right)  \nonumber
\eee
determined by the bath spectral functions $J_\phi(\omega )$, i.e. Eq.(\ref{spectra}) and Bose factor $n_B(x)=1/(\exp(x/T)-1)$. The relaxational contribution (first and second term on the r.h.s. of Eq.(\ref{chiqeq})) yields also a corresponding contribution to the spectrum $\chi''_{Q_{xy}Q_{xy}}(\omega)$.

Fig.\ref{chiq} plots the $\chi'_{Q_{xy}Q_{xy}}(T)$ versus temperature. At weak defect-phonon coupling with $\Delta_r\ll\Delta_kW$ (black full and dashed line) as well as at strong defect-phonon coupling with $\Delta_r\gg\Delta_kW$ (blue dash-dotted and doted line) one observes the $tanh$ behaviour of the resonant plus the hump due to the relaxational contribution. The latter is suppressed with increasing frequency.
We assumed in Fig.\ref{chiq} $J_{s}(2\Delta_kW)\gg J_{w,3}(\Delta_r), J_{w,1}(\Delta_r)$ and defined $\Gamma_{0s}=2\pi J_{s}(1\mbox{K})=2\pi\alpha_s (1\mbox{K})^3$ (notice that $2\Delta_kW\ll\omega_D$). The actual rates $\Gamma_{b}$ and $\Gamma_c$ are functions of $2\Delta_kW$. Accordingly the frequency $\omega_c\simeq\Gamma_{b}\simeq\Gamma_{c}$ above which the relaxational contributions vanish changes with $2\Delta_kW$.
The striking difference between weak and strong defect-phonon coupling is the temperature where the resonant response ($\tanh $ behaviour) reaches its low temperature plateau and the position of the relaxational hump.

\section{Experimental relevance}

For strong defect-phonon coupling only space diagonal, i.e. inversion symmetric, tunnelling is not suppressed. The spectrum reduces to two-states and the dielectric response is identical to a two-level tunneling system and qualitatively the same for weak defect-phonon coupling. Thus, it cannot be used to estimate the defect-phonon coupling strength. For strong defect-phonon coupling the two states are, however, 4-fold (almost) degenerate. The small splitting is determined by the suppressed tunnelling amplitudes which also sets the energy scale for the acoustic response.
In this, the acoustic response shows strikingly different behaviour at weak and strong defect-phonon coupling. Specifically, with increasing defect-phonon coupling the temperature where the resonant response ($\tanh $ behaviour) reaches its low temperature plateau shifts to lower temperatures. This, in turn, allows to determine the defect-phonon coupling by comparing dielectric and acoustic response. While at weak defect-phonon coupling both are governed by a single energy scale, i.e. $\Delta_kW$, at strong defect-phonon coupling dielectric response is governed by $\Delta_r$ but acoustic response by $\Delta_kW$.

Only for KCl doped with Li both low frequency dielectric as well as acoustic response measurements are reported at low defect concentrations. Tornow et al. report dielectric experiments at a concentration of 60 ppm \cite{Tor94} and Weiss et al. report for the same concentrations acoustic response \cite{Wei99}. In both cases resonant contributions are observed with roughly the same energy scale, i.e. $\Delta\sim 1.1$K. Weiss et al. observed additionally relaxational contributions which points clearly to a scenario for weak defect-phonon coupling. For Li defects in KCl \cite{footx} the Debye-Waller exponent is $\alpha_s\omega_D^2/\pi\simeq 7.8\cdot 10^{-2}$ in line with the experimental observation of weak defect-phonon coupling.

The Debye-Waller exponent for CN defects in KCl is $\alpha_s\omega_D^2/\pi\simeq 1.9$ which puts the system in the intermediate regime between strong and weak coupling and we expect mixed behaviour \cite{footy}. Unfortunately, only acoustic response  measurements at defect concentrations of 45 ppm (or higher) are reported \cite{Nal01,Topp02}. They exhibit clearly relaxational as well as resonant contributions. An analysis in terms of weak defect-phonon coupling successfully describes the data \cite{Nal01}. Experimental data for dielectric response in these systems would be highly interesting in order to see whether they exhibit a tunnelling energy scale identical (weak coupling scenario) or different (strong coupling scenario) from the acoustic data.

Even larger defect-phonon couplings are observed in crystals doped with OH impurities which, however, form [100] defects with 6 potential minima. Since these defects possess two tunnelling paths, of which again the geometrically shorter path is not inversion symmetric, we expect that [100] defects exhibit qualitatively the same physics as outlined in the presented theory. Ludwig et al. \cite{Lud00} observe an anomalous isotope effect when doping KCl and NaCl with OH or OD. In detail, they find that the tunneling amplitude of OH is smaller compared to the tunneling amplitude of OD although OH is lighter. They attribute this effect to strong defect - phonon coupling of the OH defect.
Furthermore, Suto and Ikezawa observed \cite{Suto84} in NaCl doped with OH that the 90$^\circ$ flip and the 180$^\circ$ flip have roughly tunneling amplitudes of the same size which points towards our proposed mechanism. The 90$^\circ$ flip tunneling amplitude is suppressed by strong defect - phonon coupling whereas the inversion symmetric 180$^\circ$ flip tunnel amplitude is unrenormalized.

All our calculations are done for a single defect in an otherwise perfect crystal, where tunneling is between local states which are unbiased with respect to each other (eight such states for the studied case of Li in KCl). However, our central result, i.e. that tunneling between inversion symmetric pairs (e.g. along the space diagonal for Li in KCl) is unaffected by strong defect-phonon coupling, carries through to the case where the local defect states are biased \cite{footpn1}. Thus, also for crystals with large defect concentration, where defect-defect interactions render finite biases, or tunneling defects in a disordered host, where disorder renders finite biases, strong defect-phonon tunneling diminishes significantly tunneling between inversion asymmetric states. This may render inversion symmetric tunneling dominant despite the larger spatial separation between its local states. Calculation of the consequences of such biases for the response functions for large defect concentrations or strongly disordered systems requires a detailed, disorder and interaction dependent, analysis and is beyond the scope of this paper.

Furthermore, it was recently suggested that tunneling two-level systems in amorphous solids result from small deviations from lattice like local structures\cite{SS13,DPRC13,DRC15}, and in that are very similar to tunneling defects in a crystal. For example, in amorphous aluminum oxide, which constitutes the barrier in Josephson junctions within superconducting qubits, it is suggested that the TLSs are single oxygens (or alternatively a small number of oxygens) tunneling between off-center positions \cite{DPRC13,DRC15}. These suggestions include tunneling between inversion symmetric (e.g. examples A and B in Fig. 1 of Ref.~\cite{DRC15}) and inversion asymmetric (e.g. example C in Fig. 1 of Ref.~\cite{DRC15}) states. Since defect-phonon coupling in amorphous solids is expected to be an order of magnitude larger than in disordered crystals \cite{GG76,BDL+85,YKMP86,Gai11}, our results suggest that the tunneling amplitude of inversion asymmetric TLSs, such as in example C in Fig. 1 of Ref.~\cite{DRC15} may be strongly diminished by defect-phonon tunneling, whereas the effect of defect-phonon interaction on the tunneling amplitude of inversion symmetric TLSs such as in examples A and B in Fig. 1 of Ref.~\cite{DRC15} is negligible. Considering a specific example of the 4-fold local degenerate state of an oxygen atom in a symmetrically pulled out aluminum cage\cite{DRC15}, our results suggest that whereas bare tunneling lifts the 4-fold degeneracy via dominant edge tunneling of the oxygen atom, in the presence of strong defect-phonon coupling the diagonal tunneling will dominate, and two 2-fold degenerate states will result (two 3-fold or 4-fold degenerate states in three dimensions, depending on the off-center direction of the localized states). Similar considerations can be applied also to quantum bits made of single crystal ${\rm Al_2O_3}$ tunnel barriers \cite{OCK+06} where tunneling TLSs may be attributed to oxygen atoms out of crystalline position at the interface layer.

The weak and strong interactions of inversion symmetric and inversion asymmetric tunneling TLSs with phonons (strain) is shown in this paper to affect significantly the TLSs' tunneling amplitude. At the same time, it affects also the typical bias energy of TLSs in strongly disordered systems, rendering a dominance of inversion symmetric TLSs and scarcity of inversion asymmetric TLSs in the single TLS density of states at low energies \cite{SS13,CBS14,CGBS13}.
Recently it was shown that Ramsey and Echo decoherence of high energy TLSs in superconducting Josephson junctions \cite{LBM+16} can be explained by the presence of thermal, weakly and strongly interacting TLSs \cite{MSSS16}, as are given by the two-TLS model\cite{SS13}. It would thus be of interest to construct a detailed picture of the tunneling entities in specific amorphous systems, and study their properties along the lines of the present work.

\section{Conclusions}

We have discussed the influence of strong phonon coupling on the dynamics of [111] tunneling defects, as for example, KCl doped with Li or CN impurities. We have specifically determined the dielectric and acoustic response. We have shown that phonon dressing of tunnelling only suppresses tunnelling along paths which are not inversion symmetric. Since the [111] defect exhibits a geometrically subdominant tunnel path along a space diagonal (and thus an inversion symmetric path), this tunnelling dominates for strong defect phonon coupling when phonon dressing suppresses all other tunnelling paths. This complements the paradigm valid for two-state systems that a defect strongly coupled to phonons cannot exhibit tunnelling. Further, we have shown that assuming strong defect-phonon coupling dielectric and acoustic response are governed by different tunnel couplings in contrast to the weak coupling case where only one tunnel coupling dominates. This results in clear qualitative differences which allow easy experimental verification. Comparing our results with available experimental data we find that Li impurities are only weakly coupled to phonons but CN impurities are more strongly coupled putting this case in an intermediate regime. We propose to carefully study dielectric response in these systems to fully characterize the defect-phonon coupling.

Strong defect-phonon coupling provides a mechanism for the preference of locally inversion symmetric two level systems. A similar preference was observed in disordered solids, therein originating from strong elastic interactions between tunneling entities. Thus, our results might help to shed light on the microscopic nature of the tunnelling systems in disordered solids responsible for their universal low temperatures properties dominated by inversion symmetric tunnelling states \cite{SS13,Gai11} as well as on the individual tunneling two-level defects observed in superconducting qubits \cite{Bar14, Mue15, LBM+16}.

\section*{Acknowledgements}

P.N. acknowledges financial support by the Deutsche Forschungsgemeinschft project NA394/2-1. M.S. acknowledges financial support by the Israeli Science Foundation (Grant No. 821/14) and by the German-Israeli Foundation (GIF Grant No. 1183/2011).

\appendix

\section{Spectral functions} \label{app1}

The bath spectral functions are determined via the defect phonon coupling, Eq. (\ref{defphoncoup}):
\[
J_\phi (\omega) = \pi\sum_{{\bf k},\alpha}
\frac{\left| \lambda_\phi ({\bf k},\alpha) \right|^2}{\omega_{{\bf k}\alpha}}
\delta(\omega-\omega_{{\bf k}\alpha})
= \alpha_\phi \omega^{x_\phi} e^{-\frac{\omega}{\omega_D}}
\]
with equal $\alpha_s$ and $x_s=3$ for $\phi=(s,xy)$, $(s,xz)$, $(s,yz)$ and
$\alpha_{w,3}$ and $x_{w,3}=5$ for $\phi=(w,3)$ and $\alpha_{w,1}$ and
$x_{w,1}=5$ for $\phi=(w,1,x)$, $(w,1,y)$ and $(w,1,z)$.

Notice that all terms
\bee
\sum_{{\bf k},\alpha}
  \frac{\lambda_{s,jl}({\bf k},\alpha) \lambda_{s,nm}^\dag({\bf k},\alpha)}
  {\omega_{{\bf k}\alpha}} &=& 0 \quad\mbox{for}\quad  (jl)\not= (nm) \nonumber\\
\sum_{{\bf k},\alpha}
  \frac{\lambda_{w,1,j}({\bf k},\alpha) \lambda_{w,1,l}^\dag({\bf k},\alpha)}
  {\omega_{{\bf k}\alpha}} &=& 0 \quad\mbox{for}\quad  j\not= l \nonumber
\eee
vanish, which greatly reduces the number of spectral functions.

We assume a Debye spectrum of phonons with linear dispersion up to the Debye frequency $\omega_D$ with an exponential cut-off function. For the coupling strengths we find $\alpha_{w3}\simeq \alpha_{w1} \simeq \alpha_s\cdot(d/v)^2$ with speed of sound $v$ which results in $J_{w,3}(\omega)\simeq J_{w,1}(\omega) \ll J_{s}(\omega)$ for all modes since their wavelength $\lambda\gtrsim \lambda_D=v\cdot(2\pi/\omega_D)\gg d$. Thus the dominant defect-phonon coupling is $W_s$.

Higher orders of the multipole expansion for the interaction of defect and lattice distortions \cite{Schechter08} result in further contributions, formally similar to $W_{w,3}$ and $W_{w,1}$ with similar coupling strengths (and weaker terms). Including them effectively changes the coupling strength $\alpha_{w3}$ and $\alpha_{w1}$ by factors of order 1.

The coupling strength $\alpha_s$ can be related to material properties \cite{Nal01}, i.e.
\be \label{alphas}
\alpha_s = \halb \sum_\alpha f_\alpha \frac{\gamma^2_\alpha}{v^5_\alpha} \frac{1}{2\pi^2 \rho \hbar}
\ee
with elastic moment $\gamma_\alpha$, speed of sound $v_\alpha$, geometric factors $f_\alpha\sim O(1)$ \cite{Nal01} for modes with polarisation $\alpha$ and mass density $\rho$ of the host crystal.

\section{Polaron transformation}\label{app2}

Using the Polaron transformation (\ref{polartrafo}) onto the Hamiltonian (\ref{hamges}) leads to the transformed Hamiltonian (\ref{hamgestraf}). In detail, we obtain the following:

\paragraph*{Defect-phonon coupling}
By definition, the dominant defect-phonon coupling $W_s$ vanishes since it is incorporated into the shifted oscillator coordinates. The subdominant defect-phonon couplings $W_{w,1}$ and $W_{w,3}$ are transformed to $\tilde{W}_{w,1}(q_{{\bf k}\alpha}) = W_{w,1}(q_{{\bf k}\alpha}+F_{{\bf k}\alpha})$
and $\tilde{W}_{w,3}(q_{{\bf k}\alpha}) = W_{w,3}(q_{{\bf k}\alpha}+F_{{\bf k}\alpha})$. Thus, new terms of the form of, for example,
$A_{z00} \sum_{{\bf k}\alpha} \lambda_{w,1,x}({{\bf k},\alpha}) F_{{\bf k}\alpha}$,
are generated. Since no phonon operator is involved anymore, the sum can be performed and we observe that all these terms vanish for symmetry reasons. Thus,
$\tilde{W}_{w,1}=W_{w,1}$ and $\tilde{W}_{w,3}=W_{w,3}$. Thus, $T^\dagger (W_s + W_{w,1} + W_{w,3}) T = W_{w,1} + W_{w,3}$.

\paragraph*{System part}
Supprisingly, the Polaron transformation leaves space diagonal tunnelling, which inverts the impurity position, $T^\dag\cdot A_{xxx} \cdot T =  A_{xxx}$ unmodified. Edge as well as face diagonal tunnelling are, however, modified:
\bee
T^\dag\cdot A_{0xx} \cdot T &=& \cos f_{xz} \left\{ A_{0xx} \cos f_{xy}  - A_{zyx} \sin f_{xy} \right\} \nonumber \\
&& -\sin f_{xz} \left\{ A_{zxy} \cos f_{xy}- A_{0yy} \sin f_{xy} \right\}\nonumber \\
T^\dag\cdot A_{x00} \cdot T &=& \cos f_{xz} \left\{ A_{x00} \cos f_{xy}  - A_{yz0} \sin f_{xy} \right\} \nonumber \\
&& -\sin f_{xz} \left\{ A_{y0z} \cos f_{xy}+ A_{xzz} \sin f_{xy} \right\} \nonumber
\eee
and $A_{x0x}$, $A_{xx0}$, $A_{0x0}$ and $A_{00x}$ accordingly (using $f_{jl}=\sum_{{\bf k}\alpha}f_{jl,{{\bf k}\alpha}}$). These modified tunnelling operators exhibit now defect as well as phonon operators. To proceed with standard system-bath methodology \cite{SpiBoLe1987, Wue98}, we split these terms into a phonon averaged part and a fluctuating part, which results in
\[
T^\dagger H_S T = \tilde{H}_S + \tilde{W}_s
\]
with $\tilde{H}_S=\langle T^\dagger H_S T \rangle_{\rm phon}$ given in Eq. (\ref{hsystraf}) where the phonons are integrated out. The Debye-Waller factor $W=\langle \cos f_{jl}\cos f_{jk} \rangle_{\rm phon}$ for $j\not= l\not= k$.

The fluctuating part constitutes
%
\bee
\tilde{W}_s &=& T^\dagger H_S T \,- \langle T^\dagger H_S T \rangle_{\rm phon} \nonumber \\
&=& -\frac{\Delta_k}{2} \Big\{ A_{x00} (\cos f_{xz} \cos f_{xy} - W)   - A_{yz0} \cos f_{xz} \sin f_{xy}
 \nonumber  \\ && \hspace*{4.5cm}  - A_{y0z} \cos f_{xy} \sin f_{xz} - A_{xzz} \sin f_{xy} \sin f_{xz}  \nonumber \\
&& \quad \qquad  +A_{0x0} (\cos f_{xy} \cos f_{yz} - W)   - A_{zy0} \cos f_{yz} \sin f_{xy}
\nonumber \\ && \hspace*{4.5cm}   - A_{0yz} \cos f_{xy} \sin f_{yz} - A_{zxz} \sin f_{xy} \sin f_{yz} \nonumber \\
&& \quad \qquad  +A_{00x} (\cos f_{xz} \cos f_{yz} - W)   - A_{z0y} \cos f_{yz} \sin f_{xz}
\nonumber \\ && \hspace*{4.5cm}    - A_{0zy} \cos f_{xz} \sin f_{yz} - A_{zzx} \sin f_{xz} \sin f_{yz} \Big\} \nonumber \\
&& -\frac{\Delta_f}{2} \Big\{ A_{0xx} (\cos f_{xz} \cos f_{xy} - W)  - A_{zyx} \cos f_{xz} \sin f_{xy}
\nonumber \\ && \hspace*{4.5cm}    - A_{zxy} \cos f_{xy} \sin f_{xz} - A_{0yy} \sin f_{xy} \sin f_{xz} \nonumber \\
&& \quad \qquad +A_{x0x} (\cos f_{xy} \cos f_{yz} - W)  - A_{yzx} \cos f_{yz} \sin f_{xy}
\nonumber \\ && \hspace*{4.5cm}  - A_{xzy} \cos f_{xy} \sin f_{yz} - A_{y0y} \sin f_{xy} \sin f_{yz} \nonumber \\
&& \quad \qquad +A_{xx0} (\cos f_{xz} \cos f_{yz} - W)  - A_{yxz} \cos f_{yz} \sin f_{xz}
\nonumber \\ && \hspace*{4.5cm}   - A_{xyz} \cos f_{xz} \sin f_{yz} - A_{yy0} \sin f_{xz} \sin f_{yz} \Big\} \nonumber
\eee

\section{RESPET}\label{app3}

We want to determine the dynamics of the defect, i.e. with Hamiltonian $\tilde{H}_S$, under the influence of the phonons with defect-phonon couplings $\tilde{W}_s + W_{w,1}+W_{w,3}$. For this, we treat the dynamics due to the full Hamiltonian (\ref{hamgestraf}) within an open quantum system framework \cite{WeissBuch} and employ explicitly the RESPET method \cite{Horner, Nal02, Nal10a, Nal14}.

Therein we do not determine the full time dependent statistical operator $R(t)$ of defect plus phonons but the reduced statistical operator of the defect $\rho(t)=\langle R(t)\rangle_{\rm phon}$ integrating out all phonon degrees of freedom. The time evolution is described by a time evolution superoperator $\U(t,t_0)R(t_0)=R(t)$ which obeys the von-Neumann equation
\[
\partial_t \U(t,t_0)=-i[\tilde{H},\U(t,t_0)]=\L\U(t,t_0)
\]
thereby defining the Liouville operator $\L$. In analogy, we can define the Liouvillians $\L_S$, $\L_W$ and $\L_0$ for the respective Hamiltonians $\tilde{H}_S$,  $\tilde{W}_{dp}=\tilde{W}_{s}+W_{w,1}+W_{w,3}$ and $H_0=\tilde{H}-\tilde{W}_{dp}$ and also corresponding time evolution operators. The full time evolution can be expressed in the form of a Dyson equation. Integrating out the phonon degrees of freedom results then in the integral equation
\be\label{Ueff} \U_{\rm eff}(t,t_0) = \U_S(t,t_0)  +\int_{t_0}^t ds\int_{t_0}^s ds' U_S(t,s) \M(s,s') \U_{\rm eff}(s',t_0)
\ee
for the effective time evolution superoperator $\U_{\rm eff}(t,t_0)$ which fulfils $\rho(t)=\U_{\rm eff}(t,t_0)\rho(t_0)$.  Herein, $\U_S(t,t_0)=\exp(\L_S(t-t_0))$ is the time evolution superoperator of the isolated defect. In order to obtain a simple representation for the memory kernel $\M(s,s')$ we employ a lowest order Born-Markov approximation in the defect-phonon coupling, reasoning that after the Polaron transformation all remaining defect-phonon couplings are weak. Note that this approach leads for a two-state problem to identical results as NIBA \cite{SpiBoLe1987, Dekker, Wue98}. We obtain
\[
 \M(s,s')= \mbox{Tr}_B\left\{ \L_W\U_0(s,s')\L_W \U_0(s',t_0)\rho_B(t_0) \right\}
\]
where the initial statistical operator is assumed to be factorized, i.e. $R(t_0)=\rho(t_0)\otimes\rho_B(t_0)$ with the initial phonon statistical operator $\rho_B=\exp(-H_B/\kb T)$  and $H_B$ the Hamiltonian of the free phonons.

With the Laplace transformation defined as
$f(z)=i\int_0^\infty dt e^{izt}f(t)$
and
$f(t)=\frac{1}{2\pi i}\int_{-\infty}^\infty dz e^{-izt}f(z)$
the Dyson equation is readily solved by
\be \U_{\rm eff}(z) \,=\, \left( \U^{-1}_S(z) \,+ \M(z) \right)^{-1}
\ee
In order to obtain the time dependence we need to Laplace back transform $\U_{\rm eff}(z)$. Thereby, we focus solely on the damping rates resulting from $\M(z)$ which are given by its imaginary part \cite{SpiBoLe1987, WeissBuch, Nal10a}. Furthermore, we assume that the memory kernel only weakly influences the poles in $\U_{\rm eff}(z)$ and thus analyse $\M(z)$ at the poles of the unperturbed system \cite{SpiBoLe1987, WeissBuch, Nal10a}, i.e. $\U_S(z)$.

The effective time evolution superoperator $\U_{\rm eff}(t,t_0)$ then allows to determine the correlation functions
\[
C_{AB}(t) = \halb\langle A(t)B+BA(t) \rangle
          = \mbox{Tr}\{ {\cal A}U_{\rm eff}(t,t_0){\cal B} \rho(t_0) \}
\]
where we define the action of ${\cal A}$ on an operator $O$ as ${\cal A}O=\halb(AO + OA)$.


\section*{Bibliography}

\begin{thebibliography}{99}

\bibitem{AHV72} P. W. Anderson, B. I. Halperin, and C. M. Varma, {\it Phil. Mag.} {\bf 25}, 1 (1972).

\bibitem{Phi72} W. A. Phillips, {\it J. Low Temp. Phys.} {\bf 7}, 351 (1972).

\bibitem{ZP71} R. C. Zeller and R. O. Pohl, {\it Phys. Rev. B} {\bf 4}, 2029 (1971).

\bibitem{FA86} J. J. Freeman and A. C. Anderson, {\it Phys. Rev. B} {\bf 34}, 5684 (1986).

\bibitem{HR86} S. Hunklinger and A. K. Raychaudhuri, {\it Prog. Low. Temp. Phys.} {\bf 9}, 265 (1986).

\bibitem{Phi87} W. A. Phillips, Rep. Prog. Phys. {\bf 50}, 1657 (1987).

\bibitem{Yu88} C.C. Yu and A. J. Leggett, Comments on Condensed Matter Physics {\bf 14}, 231 (1988).

\bibitem{PLT02} R. O. Pohl, X. Liu, and E. Thompson,
{\it Rev. Mod. Phys.} {\bf 74}, 991 (2002).

\bibitem{Ramos2014} T. Pérez-Castañeda, C. Rodríguez-Tinoco, J. Rodríguez-Viejo and M. A. Ramos, Proceedings of the National Academy of Sciences {\bf 111}, 11275 - 11280 (2014).

\bibitem{Gao08} J. Gao et al., Appl. Phys. Lett. {\bf 92}, 152505 (2008).

\bibitem{Quint14} C. M. Quintana et al., Appl. Phys. Lett. {\bf 105}, 062601 (2014).

\bibitem{SRW04} R. W. Simmonds, K. M. Lang, D. A. Hite, S. Nam, D. P. Pappas, and J. M. Martinis, Phys. Rev. Lett. {\bf 93}, 077003 (2004).

\bibitem{Bar14} R. Barends et al., Nature {\bf 508}, 500 (2014).

\bibitem{Mue15} C. M\"uller, J. Lisenfeld, A. Shnirman, and S. Poletto, Phys. Rev. B {\bf 92}, 035442 (2015).

\bibitem{LBM+16} J. Lisenfeld, A. Bilmes, S. Matityahu, S. Zanker, M. Marthaler, M. Schechter, G. Sch\"on, A. Shnirman, G. Weiss, and A.V. Ustinov, Sci. Rep. {\bf 6}, 23786 (2016).

\bibitem{Nara} V. Narayanamurti and R.O. Pohl, Rev. Mod. Phys. {\bf 42}, 201 (1970).

\bibitem{Aloisbuch} A. W\"urger, From Coherent Tunneling to Relaxation, Springer Tracts in Modern Physics {\bf 135}, Springer Heidelberg (1996).

\bibitem{Lud04} S. Ludwig and C. Enss, J. Low Temp. Phys. {\bf 137}, 371 (2004).

\bibitem{HKL90} U. T. H\"{o}chli, K. Knorr, and A. Loidl, Advances in Physics, {\bf 39}, 405 (1990).

\bibitem{YKMP86} J. J. De Yoreo, W. Knaak, M. Meissner, and R. O. Pohl, Phys. Rev. B {\bf 34}, 8828 (1986).

\bibitem{Gomez} M. Gomez, S.P. Bowen, and J.A. Krumhansl, Phys. Rev. B {\bf 153}, 1009 (1967).

\bibitem{Nal97} P. Nalbach and O. Terzidis, J. Phys.: Condens. Matter {\bf 9}, 8561 (1997).

\bibitem{Nal01} P. Nalbach, O. Terzidis, K.A. Topp, and A. W\"urger, J. Phys.: Condens. Matter {\bf 13}, 1467-83 (2001).

\bibitem{SS13} M. Schechter and P. C. E. Stamp, Phys. Rev. B {\bf 88}, 174202 (2013).

\bibitem{CBS14} A. Churkin, D. Barash and M. Schechter, Phys. Rev. B {\bf 89}, 104202 (2014).

\bibitem{Gai11} A. Gaita-Arino and M. Schechter, Phys. Rev. Lett. {\bf 107}, 105504 (2011).

\bibitem{SSMM05} A. Shnirman, G. Schon, I. Martin and Y. Makhlin, Phys. Rev. Lett. {\bf 94}, 127002 (2005).

\bibitem{MJM05} J. M. Martinis, K. B. Cooper, R. McDermott, M. Steffen, M. Ansmann, K. D. Osborn, K. Cicak, S. Oh, D. P. Pappas, R. W. Simmonds, and C. C. Yu, Phys. Rev. Lett. {\bf 95}, 210503 (2005).

\bibitem{Nems08} C. Seoánez, F. Guinea, and A. H. Castro Neto, Phys. Rev. B {\bf 77}, 125107 (2008).

\bibitem{Nems09} Laura G. Remus, Miles P. Blencowe, and Yukihiro Tanaka, Phys. Rev. B {\bf 80}, 174103 (2009).

\bibitem{Nems10} A. Venkatesan, K. J. Lulla, M. J. Patton, A. D. Armour, C. J. Mellor, and J. R. Owers-Bradley, Phys. Rev. B {\bf 81}, 073410 (2010).

\bibitem{Bey75} H. U. Beyeler, Phys. Rev. B {\bf 11}, 3078 (1977).

\bibitem{Kapp74} S. Kapphan, J. Phys. Chem. Solids 35 , 621 (1974).

\bibitem{Lud00} S. Ludwig, A. Brederek, C. Enss, C. P. An, and F. Luty, Phys. Rev. Lett. {\bf 85}, 5591 (2000).

\bibitem{foot5} Higher order terms can safely be disregarded since the defect size $d$ is typically smaller than the lattice constant $a$.

\bibitem{foot1} Notice that the complex $i$ is necessary in order to ensure ${\bf u}({\bf r})$ to be Hermitian since ${\boldsymbol \xi}_{-{\bf k}\alpha}=-{\boldsymbol \xi}_{{\bf k}\alpha}$.

\bibitem{SpiBoLe1987} A.J. Leggett, S. Chakravarty, A.T. Dorsey, M.P.A. Fisher, A. Garg, and W. Zwerger, Rev. Mod. Phys. 59 (1987) 1.

\bibitem{WeissBuch} U. Weiss, Quantum Dissipative Systems, 3rd ed., World Scientific, Singapore, 2007.

\bibitem{Dekker} H. Dekker, Phys. Rev. A {\bf 35}, 1436 (1987).

\bibitem{Wue98} A. W\"urger, Phys. Rev. B {\bf 57}, 347 (1998).

\bibitem{Horner} H. Horner, Eur. Phys. J. B {\bf 18}, 453 - 458 (2000).

\bibitem{Nal02} P. Nalbach, Phys. Rev. B {\bf 66}, 134107 (2002).

\bibitem{Nal10a} P. Nalbach and M. Thorwart, J. Chem. Phys. {\bf 132}, 194111 (2010).

\bibitem{Nal14} P. Nalbach, Phys. Rev. A {\bf 90}, 042112 (2014).



\bibitem{foot4} When $\Delta_kW-\Delta_r\ll\hbar\omega$ is fulfilled, we observe relaxation behaviour with a finite loss contribution and strong frequency dependence in both, loss and the real part of the susceptibility. For experimental frequencies in the kHz regime such a fine tuned phonon renormalization of the edge tunneling seems unrealistic.

\bibitem{foot3} We omit the spatial indices at the spectral functions since $J_{s,xy}=J_{s,xz}=J_{s,yz}$ and $J_{w,1,x}=J_{w,1,y}=J_{w,1,z}$ due to symmetry.

\bibitem{Tor94} M. Tornow, R. Weis, G. Weiss, C. Enss, and S. Hunklinger, Physica B {\bf 194-196}, 1063 (1994).

\bibitem{Wei99} G. Weiss, M. H\"ubner, and C. Enss, Physica B {\bf 263-264}, 388 (1999).

\bibitem{footx} The elastic moment is $\gamma_t=0.04$eV (from Ref.\cite{Hueb94}), the speed of transversal sound for KCl is $v_t=1.7$km/s (from Ref.\cite{Nal01}), its mass density $\rho=1.989$g/cm$^3$ (from Ref.\cite{Nal01}), the Debye temperature is $\hbar\omega_D/\kb=230$K for KCl (from Ref.\cite{Tech}) and the geometry factor \cite{Nal01} is $f_t=1/5$. With these parameters the DEbye-Waller exponent Eq.(\ref{alphas}) can be determined keeping in mind that the longitudinal speed of sound is larger and the transversal one and thus contributions from longitudinal phonon branches are negligible in Eq.(\ref{alphas}).

\bibitem{footy} We used a tunneling amplitude $\Delta\sim 1.55$K and an elastic moment $\gamma_t=0.192$eV \cite{Nal01}.

\bibitem{Topp02} K. A. Topp and R. O. Pohl, Phys. Rev. B {\bf 66}, 064204 (2002).

\bibitem{Suto84} S. Suto and M. Ikezawa, J. Phys. Soc. Jpn. 53 , 438 (1984).

\bibitem{footpn1} Biases in an [111]-defect are modelled by operators $A_{\alpha\beta\gamma}$ with $\alpha$, $\beta$ and $\gamma$ equal to $0$ or $z$. Thus, the Polaron transformation $T$ commutes with the biases and accordingly leaves them unmodified.

\bibitem{DPRC13} T. C. DuBois, M. C. Per, S. P. Russo , J. H. Cole, Phys. Rev. Lett. {\bf 110}, 077002 (2013).

\bibitem{DRC15} T. C. DuBois, S. P. Russo , J. H. Cole, New J. Phys. {\bf 17}, 023017 (2015).

\bibitem{OCK+06} S. Oh, K. Cicak, J. S. Kline, M. A. Sillanpaa, K. D. Osborn, J. D. Whittaker, R. W. Simmonds, D. P. Pappas, Phys. Rev. B {\bf 74}, 100502 (2006).

\bibitem{GG76} B. Golding and J. E. Graebner, Phys. Rev. Lett. {\bf 74}, 852 (1976).

\bibitem{BDL+85} J. F. Berret P. Doussineau, A. Levelut, M. Meissner, W. Schon, Phys. Rev. Lett. {\bf 55}, 2013 (1985).

\bibitem{YKMP86} J. J. De Yoreo, W. Knaak, M. Meissner, R. O. Pohl, Phys. Rev. B {\bf 34}, 8828 (1986).

\bibitem{CGBS13} A. Churkin, I. Gabdank, A. Burin, and M. Schechter, arXiv:1307.0868.

\bibitem{MSSS16} S. Matityahu,A. Shnirman,G. Sch\"on, M. Schechter, Phys. Rev. B {\bf 93}, 134208 (2016).


\bibitem{Schechter08} M. Schechter and P. C. E. Stamp, J. Phys.: Condens. Matter {\bf 20}, 244136 (2008).

\bibitem{Hueb94} M. H\"ubner, diploma thesis (1994), Ruprechts-Karls Universit\"at Heidelberg

\bibitem{Tech} value taken from http://www.techniklexikon.net/d/debye-temperatur/debye-temperatur.htm

\end{thebibliography}
\end{document}